%% file: SSC_inOrbi_PerformanceV1.0.tex
\documentclass[proof]{pasj00}
\usepackage{lscape}
\draft
\newcommand{\Rcycle}{{5.865}}  
\newcommand{\SSCeff}{{30}}  

\begin{document}
\SetRunningHead{H. Tsunemi et al.}{SSC onboard ISS} 
\Received{2010/07/09}
\Accepted{}

\title{In Orbit Performance of the MAXI/SSC onboard the ISS }  

\author{  Hiroshi \textsc{Tsunemi}\altaffilmark{1},
	   Hiroshi \textsc{Tomida}\altaffilmark{2},
	   Haruyoshi \textsc{Katayama}\altaffilmark{3},
	   Masashi \textsc{Kimura}\altaffilmark{1}, \\
	   Arata  \textsc{Daikyuji}\altaffilmark{4},
	   Kazuhisa \textsc{Miyaguchi}\altaffilmark{5},
	   Kentaro \textsc{Maeda}\altaffilmark{5},
	   and
	   MAXI \textsc{Team}\altaffilmark{1}
 }
 \altaffiltext{1}{Department of Earth and Space Science, Graduate School of 
                  Science, Osaka University, 1-1 Machikaneyama, Toyonaka, Osaka
                  560-0043, Japan
                  }
 \email{tsunemi@ess.sci.osaka-u.ac.jp}
 \altaffiltext{2}{ISS Science Project Office, ISAS, JAXA, 2-1-1 Sengen, Tsukuba, Ibaraki 305-8505, Japan
                  }
 \altaffiltext{3}{Earth Observation Research Center, JAXA, 2-1-1, Sengen, Tsukuba, Ibaraki 305-8505, Japan}

 \altaffiltext{4}{Department of Applied Physics, Faculty of Engineering, University of Miyazaki, 1-1 Gakuen Kibana-dai
Nishi, Miyazaki, 889-2192, Japan}

 \altaffiltext{5}{Solid State Division, Hamamatsu Photonics K.K., 1126-1 Ichino, Hamamatsu 435-8558, Japan}

%

\KeyWords{instrumentation: detectors -- space vehicles: instruments -- X-ray CCDs} 

\maketitle

\begin{abstract}
We report here the in orbit performance of the CCD camera (MAXI/SSC) onboard the International Space Station (ISS).  It was commissioned in August, 2009.  This is the first all-sky survey mission employing X-ray CCDs.  It consists of 32 CCDs each of which is 1 inch square.  It is a slit camera with a field of view of 1$^\circ$.5$\times 90^\circ$ and scans the sky as the rotation of the ISS.

The CCD on the SSC is cooled down to the working temperature around $-$60$^\circ$C by the combination of the peltier cooler, a loop heat pipe and a radiator.  The standard observation mode of the CCD is in a parallel sum mode (64-binning).  The CCD functions properly although it suffers an edge glow when the Sun is near the field of view (FOV) which reduces the observation efficiency of the SSC down to about \SSCeff \%.  The performance of the CCD is continuously monitored both by the Mn-K X-rays and by the Cu-K X-rays.   

There are many sources detected, not only point sources but extended sources.  Due to the lack of the effective observation time, we need more observation time to obtain an extended emission analysis extraction process.
\\
\end{abstract}

\input{Introduction}

\input{Structure}


\input{onboardDataProcess}

\input{Calibration}

\input{ScanObservation}

\input{AllSkyMap}

\input{Conclusion}

\section*{Acknowledgments}

This work is partly supported by a Grant-in-Aid for Scientific
Research by the Ministry of Education, Culture, Sports, Science and Technology (16002004).  M. K. is supported by JSPS Research Fellowship for Young Scientists (22$\cdot$1677).

\input{Reference}

\end{document}

%% file: Introduction.tex
\section{Introduction}

The ASCA satellite (\cite{tanaka1994}) was the first successful mission that carried X-ray CCDs in photon counting mode (\cite{burke1991}).  Since then, the X-ray CCDs in photon counting mode become the standard X-ray detectors for the X-ray astronomy missions.  Chandra (\cite{weisskopf2002}) launched in 1999 carries the ACIS consisting of 10 CCD chips each of which is one inch square in imaging area.  XMM-Newton launched in 1999 carries the EPIC that consists of two CCD cameras: one is the MOS camera (\cite{turner2001}) and the other is the pn camera (\cite{struder2001}).  The MOS camera consists of 7 CCD chips each of which is 4 cm square.  The pn camera has one CCD chip that is 6\,cm square.  The Swift satellite launched in 2004 (\cite{gehrels2004}) carries one CCD chip that is identical to those of the MOS camera.  The Suzaku satellite launched in 2005 carries the XIS (\cite{koyama2007}) consisting of 4 CCD chips that are improved CCDs of the ACIS.

The MAXI (Monitor of All-sky X-ray Image, \cite{matsuoka2009}) was planned in 1998 to be launched in 2003.  There are two cameras: the Gas Slit Camera (GSC, \cite{mihara2010}) and the Solid-state Slit Camera (SSC).  The SSC was proposed to be an array of X-ray CCD onboard the MAXI.  This is the slit camera: a kind of a pin hole camera.  It carries 32 CCD chips each is one inch square.  Due to the unexpected problem of the ACIS in orbit (\cite{weisskopf2002}), we introduced a continuous function of the charge injection so that we could recover the radiation damage on the CCD.  This technique (\cite{prigozhin2008}) is added to the XIS in orbit and improves the performance of the CCD (\cite{uchiyama2009}).  Then, the MAXI was successfully launched in July, 2009 by the space shuttle, Endeavor, and was installed to the Japanese Experiment Module - Exposed Facility (JEM-EF) on the International Space Station (ISS).

We report here the basic structure, calibration and in orbit performance of the SSC.

%% file: Structure.tex
\section{SSC Structure and its thermal performance in orbit}


The SSC is installed on the GSC that is described in detail (\cite{matsuoka2009}).  There are two cameras: SSC-H and SSC-Z.  Each has 16 CCD chips and is identical to each other with the exception that the SSC-H is watching +20$^\circ$ above the horizontal direction of the ISS and the SSC-Z is watching the zenith direction of the ISS.  We aligned 16 CCD chips in 2$\times$8 array.

\subsection{CCD for the MAXI/SSC}

We have developed the CCD for X-ray detection (\cite{tsunemi2007}) under the collaboration with the Hamamatsu Photonics K.K. Table \ref{CCD_parameter} shows the structural parameters of the CCD (\cite{katayama2005}).  Since there is no storage area, it works in a full frame mode.  The X-ray entrance surface has an Al coat to prevent visible light from entering.  The Al coat on the CCD enables it to eliminate a optical blocking filter in front of the CCD.  This also makes it possible to eliminate vacuum tight body.  The CCD has a thick depletion layer by using a high resistivity wafer.   Based on the design parameters and the ground calibration (\cite{tomida2010}), we calculated the quantum efficiency (QE) that is shown in the literature (\cite{matsuoka2009}).  In the next section, we will improve the accuracy of the QE calibrated in orbit by using the Crab nebula.

\begin{table}[hbt]
\begin{center}
\caption{\sc MAXI-CCD parameters } 
\label{CCD_parameter}
\begin{tabular}{lr} \hline\hline
IA \# of pixels & $1024 \times 1024$\\
IA pixel & $24\mu{\rm m}\times 24\mu{\rm m}$\\
Type & Front-illuminated \\
\# of nodes & 1\\
wafer type & p-type\\ 
coat & Al (2000\AA )\\ 
\hline\hline
\end{tabular}
\end{center}
\end{table}
 
\subsection{Thermal performance in Orbit}

The use of CCDs for X-ray detection requires a low working temperature.  The cooling system of the MAXI/SSC consists of two parts: one is a peltier device and the other is a radiator with a loop heat pipe (LHP).



The hot side of the peltier is thermally connected to the body of the SSC.  The body of the SSC is cooled down around $-$20$^\circ$C that depends on the thermal condition of the radiator.  The radiator consists of two panels: an upper panel (Radiator Z : 0.527\,m$^2$) and a forward panel (Radiator H : 0.357\,m$^2$).  They are designed such that we can obtain the maximum area within the allocated volume of the MAXI.  Since these two panels are not in the same plane, the LHP is designed to cool down the body of the SSC as much as we can.  After  launch, the LHP started properly and is working as we expected.  The radiator temperature depends on the ISS location and its orientation, but there has been no decrease of cooling power as of the writing phase of this paper.

\begin{figure}
  \begin{center}
\includegraphics*[width=8cm, bb=0 0 634 473]{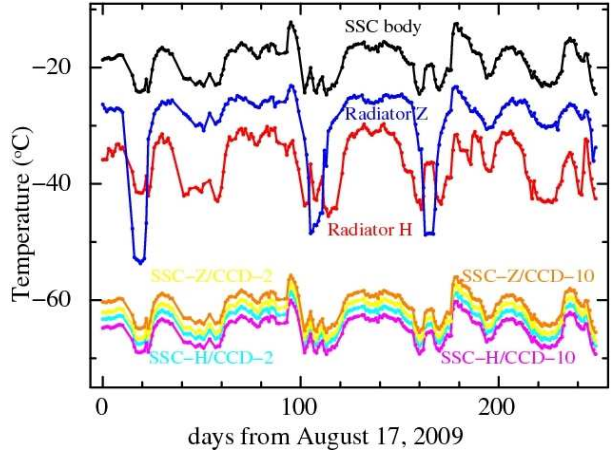}
\end{center}
  \caption{Temperature histories of the CCD, the SSC body and two panels of the radiator are shown since the launch of the MAXI/SSC.  We showed temperatures for 4 CCDs out of 32.}\label{CCD_temp}
\end{figure}

There are two means of temperature control.  One is to keep the CCD temperature constant while the other is to keep peltier current constant.  Since we select the latter mean in the present operation of the SSC, the temperature of the CCD changes as the radiator temperature changes.  Figure \ref{CCD_temp} shows the temperature history of the CCD after the launch where each point represents the one-day averaged temperature.  Two panels of the radiator show different thermal behavior whose temperature is between  $-$25$^\circ$C and $-$55$^\circ$C.  The LHP cools down the SSC body around $-$20$^\circ$C.  The peltier device cools down the CCD around $-$60$^\circ$C.  Since the peltier is running at constant current mode, the temperature difference is also constant around 45$^\circ$C depending on the individual peltier device.  The detailed spectral analysis takes into account the working condition of the SSC.

%% file: onboardDataProcess.tex
\section{Onboard Data Process}    

The data handling scheme of the MAXI/SSC is similar to that of the previous satellites.  The detailed explanation is given in the literature (\cite{tomida2010}).  There is one video chain for each camera (16 CCDs).  MAXI is designed to observe not in a pointing mode but in a scanning mode.  We have two data-taking modes: one is the full frame mode and the other is the parallel sum mode.   The full frame mode is to obtain the image of 1024$\times$1024 pixels for each CCD.  The full frame mode is only used to check the performance of the CCD since the time resolution is so bad due to the telemetry capacity.

The parallel sum mode is to add up many pixel signals to speed up the read out time.  In the standard observation of the MAXI/SSC, we add 64 rows as on-chip sum while other binning is possible by command.  As we employ a charge injection at every 64 rows, we skip the charge injected row and accumulate 63 rows for binning.  In this way, we obtain 16$\times$1024 pixels for each CCD.  Since we sequentially read 16 CCD chips, the read-out time is \Rcycle\,s that depends on the number of rows of the on-chip sum.  We should note that each CCD will have an integration time of 5.498\,s followed by a read-out time of 0.367\,s since one read-out video chain reads 16\,CCDs one by one.  This makes us possible to partly determine the incident position of the X-ray photon within the CCD.  In this way, we can measure the CTI of the CCD even in the parallel sum mode.

We employ event recognition method similar to the ASCA grade.  In the parallel sum mode, the charge spread of the signal is effectively valid only for G0 (single event), G1 (left split event), G2 (right split event) and G3 (three-pixel event).  Since we do not see the charge spread in the vertical direction, the background rejection efficiency is worse than that of the normal mode in other satellites.  We expect that the X-ray events form G0, G1 or G2 while the charged particle event forms G3.

\begin{figure}
  \begin{center}
\includegraphics*[width=6cm, bb=0 0 588 589]{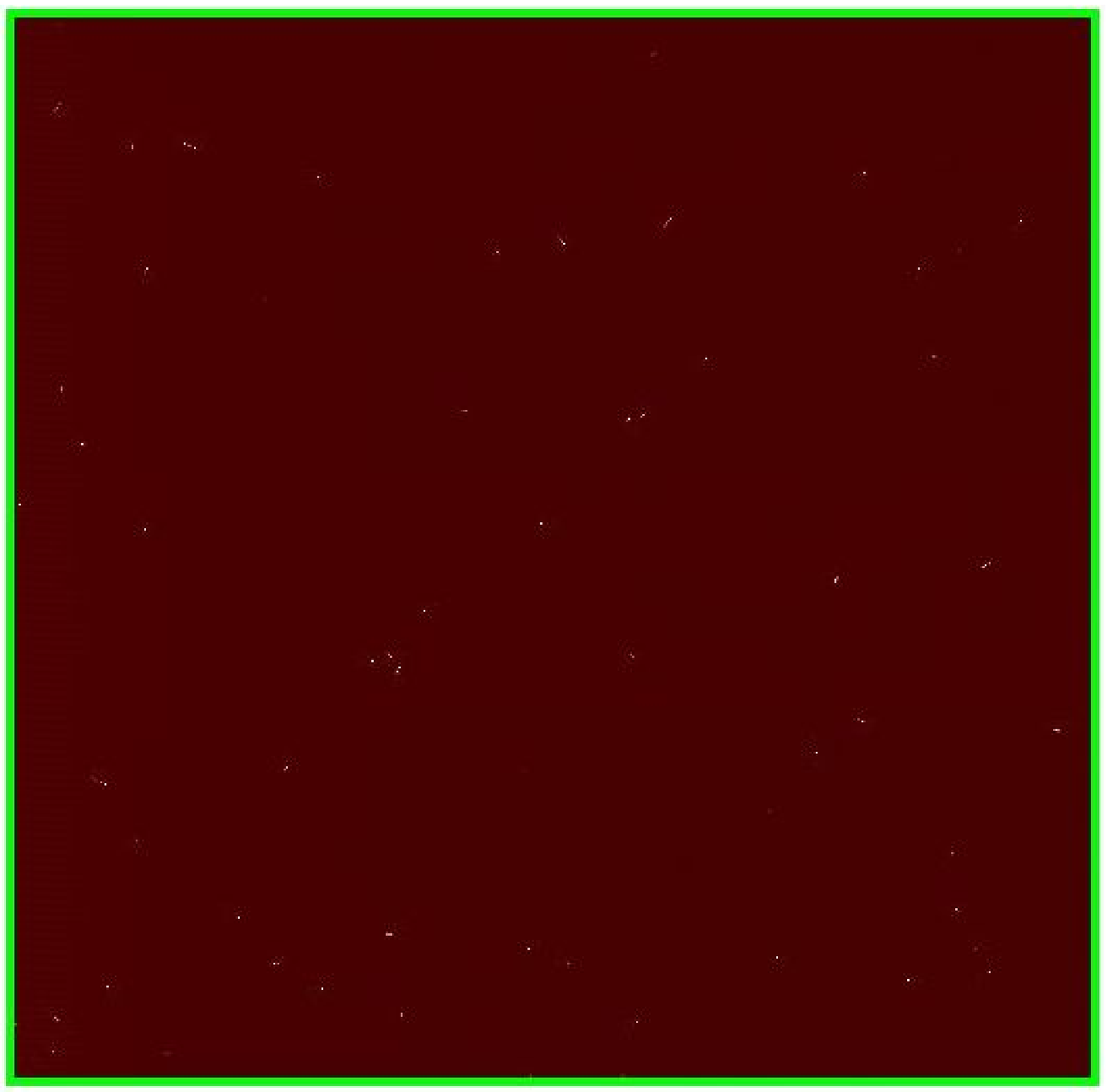}
\includegraphics*[width=6cm, bb=0 0 588 589]{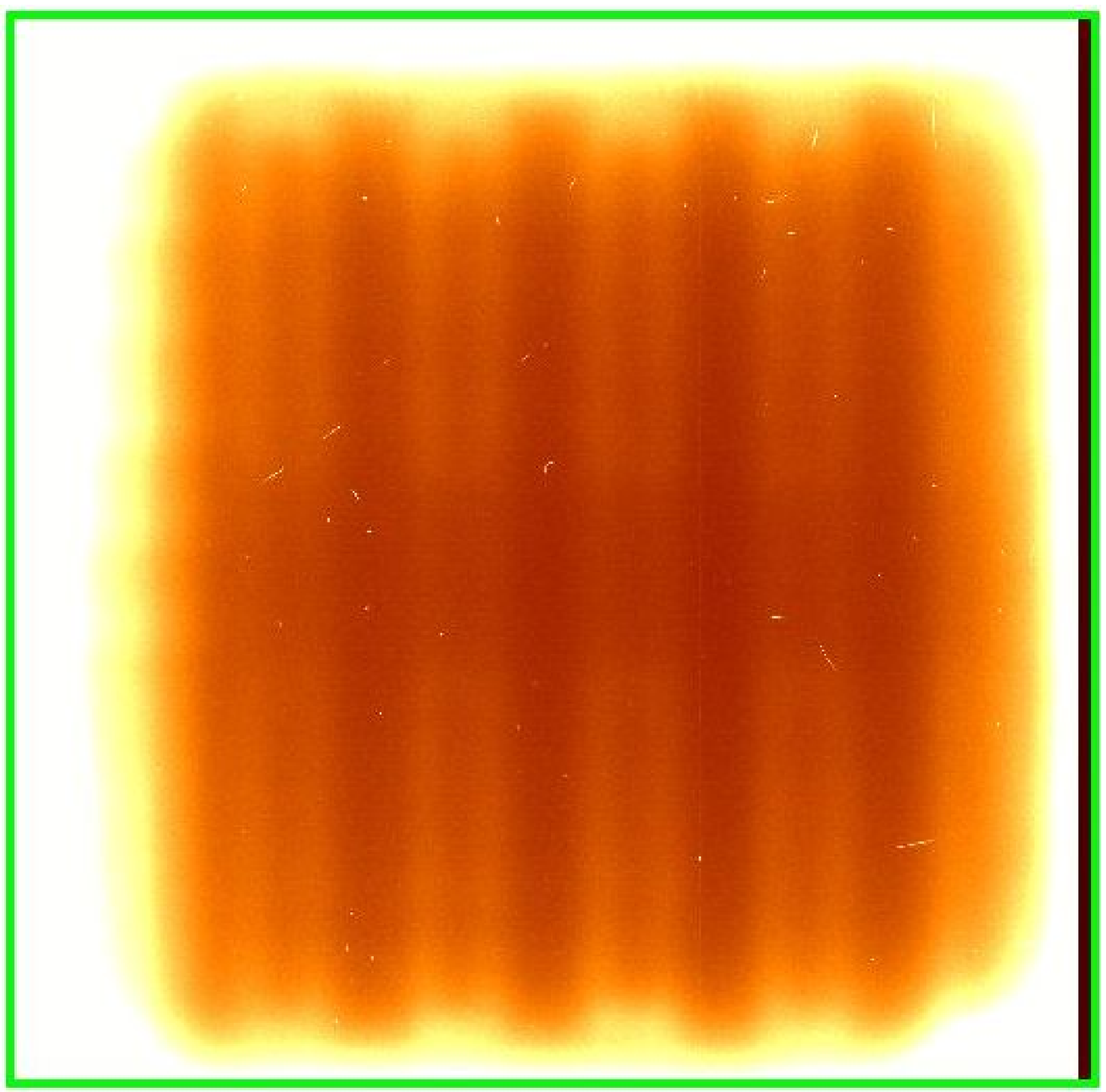}
\end{center}
  \caption{Frame images obtained at night time (left), at day time (right).  Each frame consists of 1024$\times$1024 pixels which is used for diagnostics.  In the normal observation mode, parallel sum mode (64 binning) is employed.}\label{frame_image}
\end{figure}

Figure \ref{frame_image} shows typical frame images obtained in orbit.  The night time image shows X-ray events and particle events as well as some fixed patterns due to the CCD operation.  The day time image shows overflow in the edge area of the CCD.  During the day time, direct Sun lights enter through the slit and scattered inside the collimator even if the Sun does not illuminate on the CCD.  In the central part of the CCD, the Sun light is well blocked by the Al coat.  Because the edge of the Si wafer of the CCD is left un-coated,  Sun light penetrates into the CCD.  Since the edge glow appears only in the day time observation, we think that the bright IR light enters the CCD through the edge of the CCD.  

Looking at the night time image, we can handle the parallel sum mode data as other satellites do.  The day time image indicates that pixels on the edge of the CCD show saturation in the electronics.  Therefore, only a central part of the image can be analyzed.  In the parallel sum mode in 64-binning, the effective area reduces by 70\%.  Furthermore, the background is quite different from that of the night time image.  We are still studying how to use the day time image.

%% file: Calibration.tex
\section{Calibration}   

\subsection{Performance of the CCD in orbit}

\begin{figure}
  \begin{center}
\includegraphics*[width=6cm, bb=0 0 638 995]{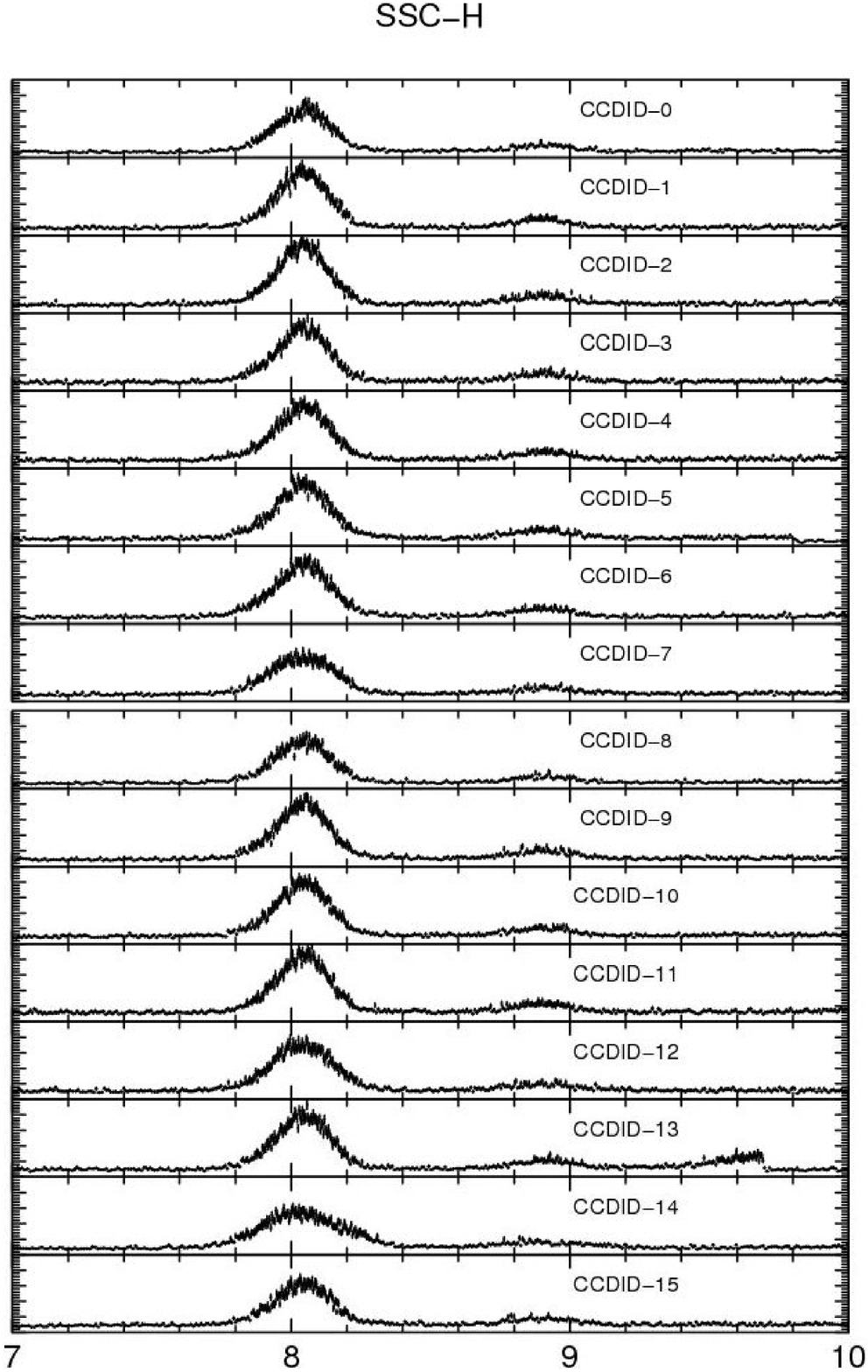}
\includegraphics*[width=6cm, bb=0 0 638 992]{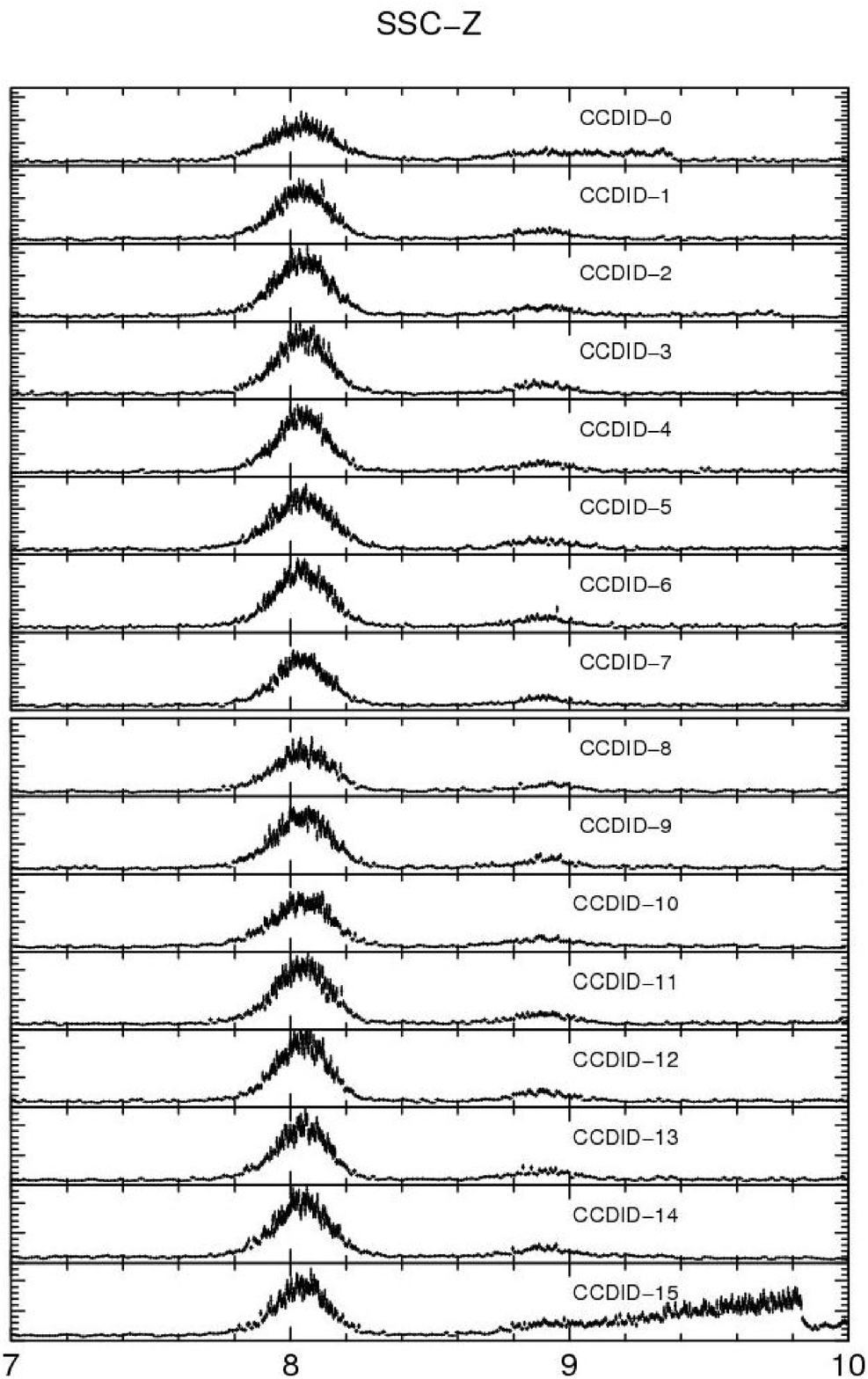}
\end{center}
  \caption{Spectra for Cu-K lines for 32 CCDs.  The effective energy ranges of some CCDs are limited below 9\,keV due to the electronics. }\label{PHA_32CCD}
\end{figure}

Radio-active sources of $^{55}$Fe are installed at the edge of the cameras inside the MAXI/SSC .  We also obtain continuous monitoring of Cu-K lines produced at the collimator by the incident particles.  Figure \ref{PHA_32CCD} shows the spectra for 32 CCDs each of which shows Cu-K lines.  By using these lines, we can continuously monitor the performance of the CCD.  The intensity of Cu-K lines is about 0.08 counts s$^{-1}$ per CCD and that of Mn-K lines is about 0.05 counts s$^{-1}$ per CCD at the time of launch.

\begin{figure}
  \begin{center}
\includegraphics*[width=8cm, bb=0 0 647 467]{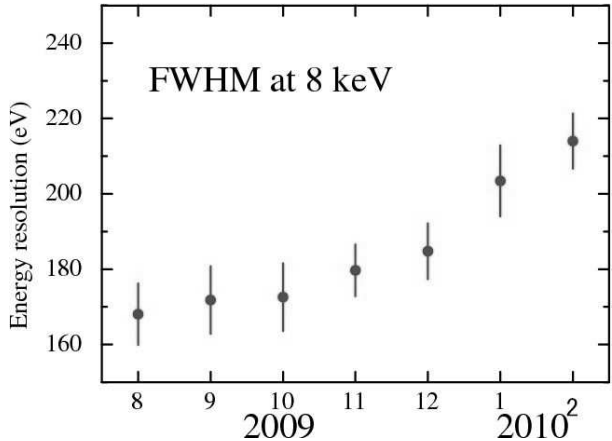}
\end{center}
  \caption{History of the energy resolution (FWHM) at the Cu-K line is shown since the launch of the MAXI/SSC. }\label{Cu_K_history}  
\end{figure}

Figure \ref{Cu_K_history} shows the history of the Cu-K peak position and its energy resolution (full width at half maximum: FWHM).  Each point represents the data for one-day accumulation time.  The FWHM of Mn-K$\alpha$ is 147\,eV and that of Cu-K$\alpha$ is 170\,eV at the time of launch.  Mn-K is irradiated onto the far side from the read-out node of 4 CCDs while Cu-K is irradiated uniformly over all the CCDs.

The ISS is in a circular orbit of altitude around 350\,km and inclination of 51$^\circ$.6.  It passes through the South Atlantic Anomaly (SAA) for 9-10 times a day.  Furthermore, it passes through high background region at high latitude.  The background is monitored by the RBM onboard the GSC (\cite{sugizaki2010}).  It shows that 15\% of the time is high background due to SAA passage.  Furthermore, it shows that the passage through high latitude also shows high background.  The background spectra for these two passages may be different whereas they are so high that the GSC turns off in observation.

The CCD employed has two characteristics for radiation-hardness.  One is a notch structure that confines the charge transfer channel to a very narrow width.  This improves a radiation-hardness by a factor of three (\cite{tsunemi2004}).  The other is a charge-injection (CI) gate through which we can continuously inject some amount of charge at every 64 rows in 64 binning mode (\cite{miyata2002}).  The CI method can partly compensate for degradation of the charge transfer inefficiency (CTI) of the CCD (\cite{tomida1997}).  We can confirm the validity of the CI by stopping the operation of the CI.  As of the writing phase of this paper (8 months after launch), we found that the decay of the Cu-K line of about 1\,\% /year without CI while that it improves to be less than 0.2\,\% /year with CI.  The decay of the FWHM at 6\,keV is estimated to be 60\,eV/year.  This value is more than we expected before launch (\cite{miyata2002}).

The Suzaku XIS employs CI in orbit (\cite{prigozhin2008}).  CI of the XIS (FI) improved the decay of the 6\,keV line from 1.6\% /year to 0.4\% /year.  It also improves the decay of FWHM at 6\,keV from 60\,eV/year to 20\,eV/year.  Therefore, the performance of the CI on the XIS is better than that of the SSC.  There are two differences between the XIS and the SSC.  One is the working temperature.  The XIS is working at -90$^\circ$C while the SSC is working at -60$^\circ$C.  The other is the background condition.  The high background passage time is 55\% for the MAXI/SSC while that is 10\% for Suzaku.  Although the effect of the background passage depends on the spectrum and its intensity, the difference between Suzaku and MAXI/SSC is due to the difference in orbit, mainly in inclination angle.  

\subsection{Background of the SSC in orbit}

The background condition is carefully monitored by the RBM and the GSC on MAXI since inclination of the ISS orbit is 51$^\circ$.6 resulting to pass at higher latitude regions than other X-ray missions.  We switch off the bias level to the CCD during the passage of the SAA (\cite{miyata2003}).  However, we switch on during the passage of high latitude regions where the background is too high to obtain good data.

\begin{figure}
  \begin{center}
\includegraphics*[width=6cm, bb=0 0 1075 1037]{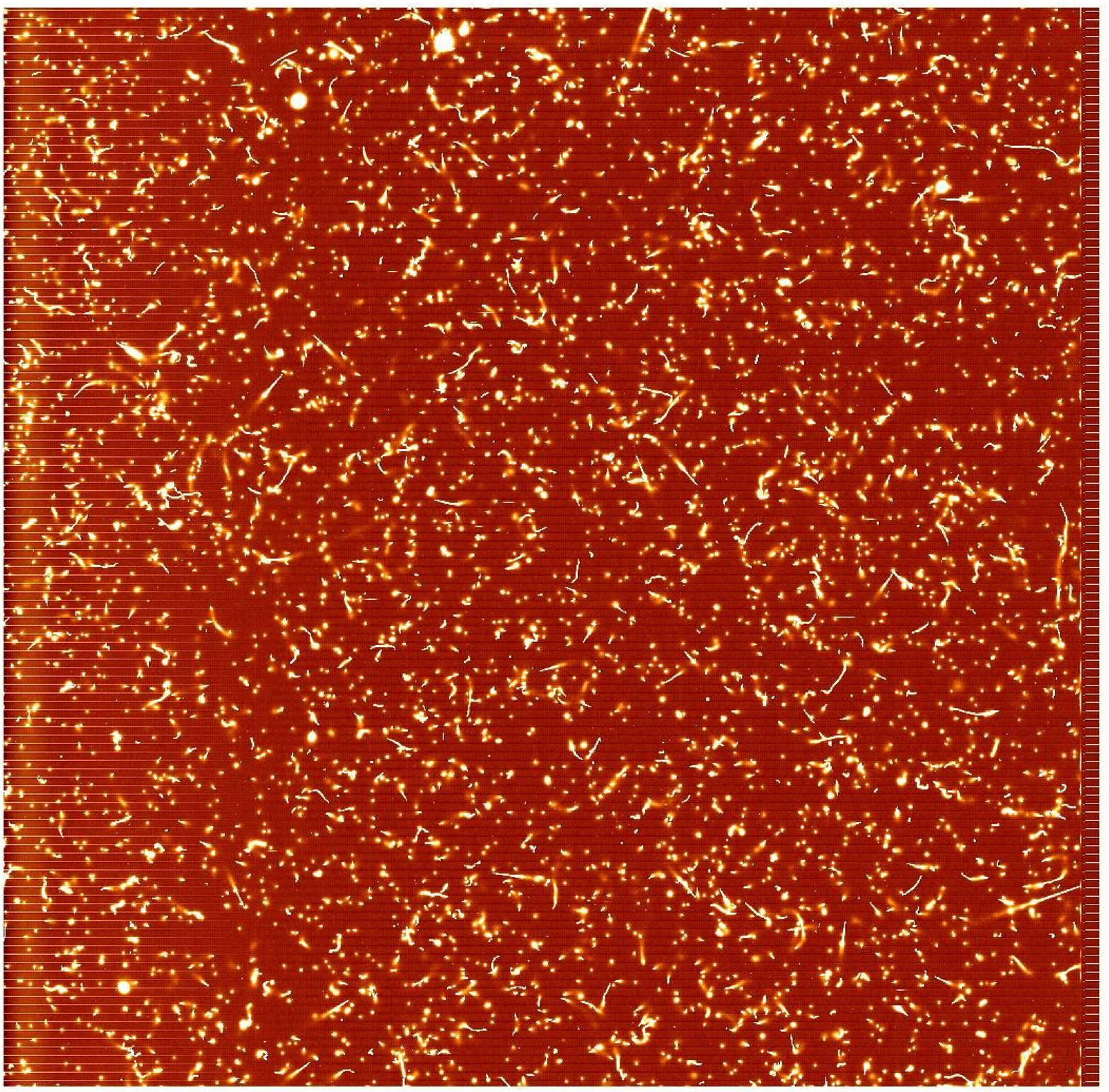}
\includegraphics*[width=6cm, bb=0 0 1075 1037]{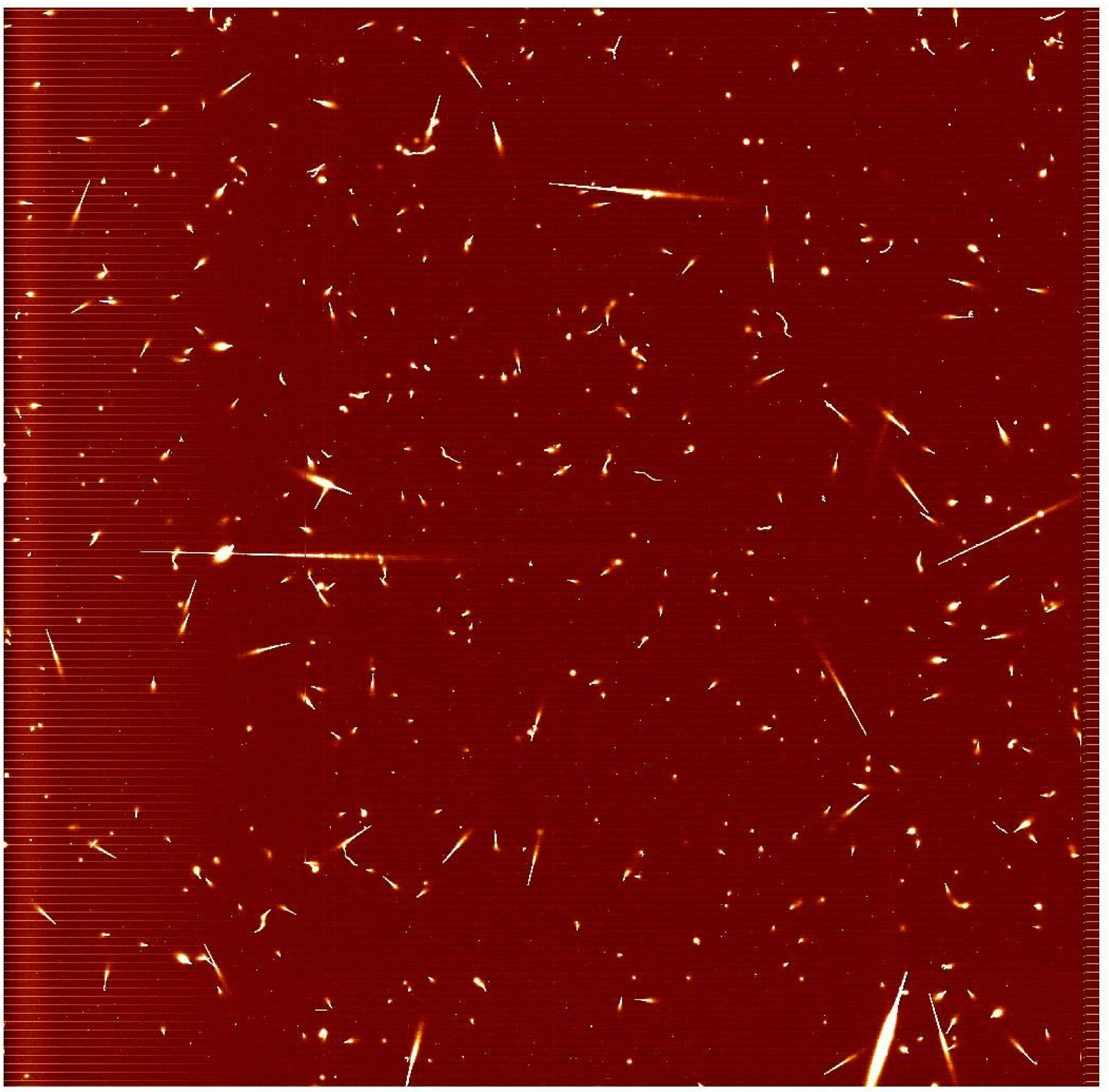}
\end{center}
  \caption{Frame images are shown at high background region.  The left image is obtained from the SSC-H and the right from the SSC-Z.}\label{frame_image_particle}
\end{figure}

Figure \ref{frame_image_particle} shows the frame image obtained near the high background region (W108$^\circ$.6, N49$^\circ$.4, cut-off-rigidity (COR) = 1.0\,GeV/c) at the same exposure time at 8:34, April 21, 2010.  We clearly see many particle events on both the SSC-H and the SSC-Z.  The SSC-H image shows many circular extended events that are generated by particles entering into the CCD along the normal to the CCD surface.  The SSC-Z image shows events of long trajectories that run within the depletion layer.  The number of particle events on the SSC-H image is about 10 times more than that of the SSC-Z.  This indicates that the cross-section against the particle flux is much bigger in SSC-H than that in SSC-Z.  Taking into account the fact that the direction of the geomagnetic field is 74$^\circ$ downward from the horizon, the normal of the SSC-Z is almost parallel to the geomagnetic field line while that of the SSC-H is almost perpendicular to it.  Since the high background comes from the trapped charged particles by the geomagnetic field, their moving direction is almost in the horizontal direction.  Therefore, they leave events of long trajectories with less flux in the SSC-Z while they leave circular extended events with more flux in the SSC-H.

\begin{figure}
  \begin{center}
\includegraphics*[width=6cm, bb=0 0 644 466]{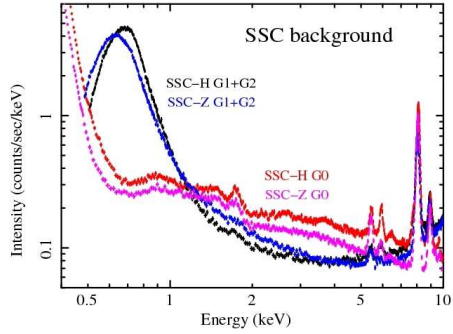}
\end{center}
  \caption{Background spectrum for G0 and G1+ G2.}\label{background_grade}  
\end{figure}

We have accumulated data during times when the background level is low.  All the events are sorted according to the grade; G0, G1, G2 and G3.  G3 is thought to be generated from particle events.  Figure \ref{background_grade} shows the spectra for G0 and G1+G2.

The G0 spectrum shows two components dividing at 0.6\,keV.  The G1+ G2 spectrum has three components dividing at 1.3\,keV and 0.7\,keV.  We found that the G1 spectrum below 0.6\,keV is produced by particles having long trajectory on the CCD.  The very end of the trajectories sometimes leave G0 events.  If they leave adjacent two pixels, they will form G1 or G2 events that extend the spectrum of G1 and G2 up to 1.3\,keV.  If they leave more than three pixels, they will be treated as particle background.  The peaks around 0.7\,keV on the G1+G2 spectra come from the event threshold (0.4\,keV).  In this way, we find that the parallel sum mode on the MAXI/SSC generates high background at low energy depending on the grade.

\begin{figure}
  \begin{center}
\includegraphics*[width=6cm, bb=0 0 644 481]{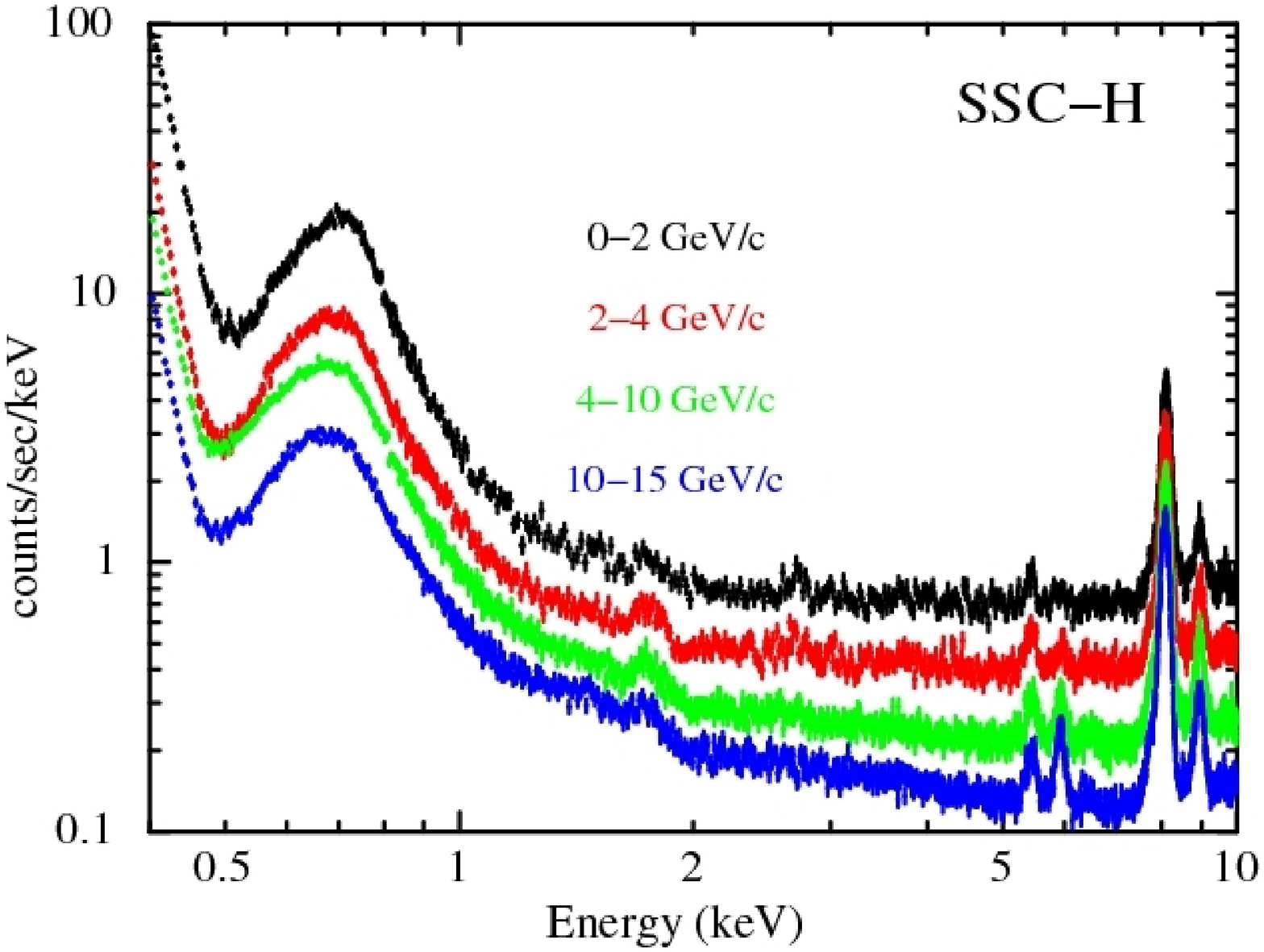}
\includegraphics*[width=6cm, bb=0 0 644 481]{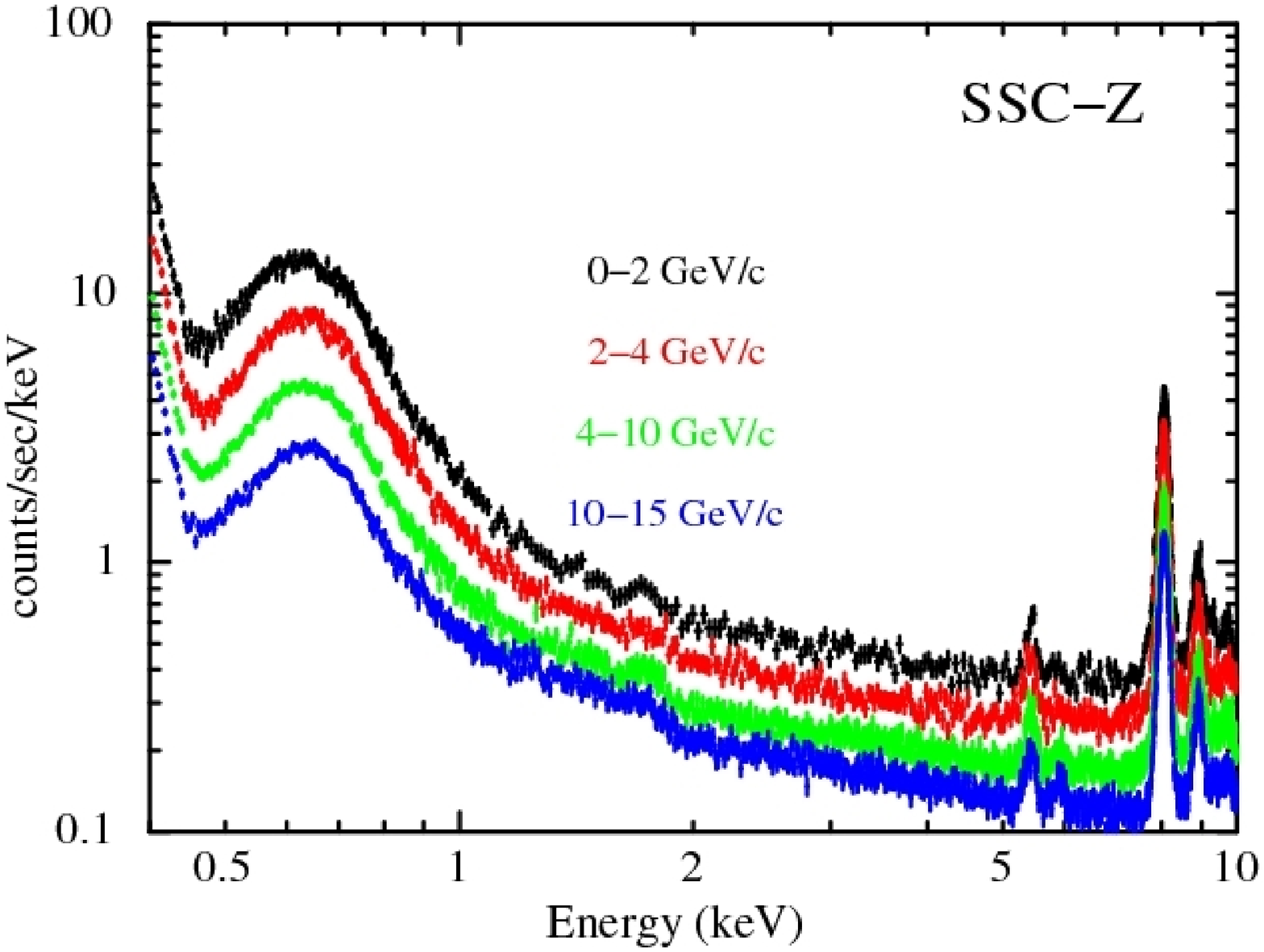}
\end{center}
  \caption{Background spectrum of G0+G1+G2 for various levels of the COR.}\label{background_COR}  
\end{figure}

Figure \ref{background_COR}  shows the background spectra of G0+G1+G2.  We clearly see the anti-correlation between the intensity and the COR. Furthermore, the spectral shape is almost constant.  Therefore, we have to sort the data by the COR.  Since we observe the sky in a scanning mode, it is relatively easy to estimate the background for point sources.  However, it is difficult to estimate the background for extended structures.  The precise analysis for extended sources require us to study the background behavior as a function of the COR. 

Hiraga et al. (2001) measured the charge spread in the CCD as a function of mean absorption length in Si.  Then, we can calculate the branching ratio for X-ray events.  For example, the Al-K X-rays generate 92\% for G0 and others for G1 and G2 that is confirmed by the ground calibration.  From figure \ref{background_grade}, we see that the fraction of G1 and G2 increases as a function of energy, particularly above 5\,keV.  This indicates that the charge spread increases as a function of energy.  Taking into account the background condition, we use only  G0 events for the energy range below the Si-K edge and G0+G1+G2 events for the energy range above it.

\subsection{Calibration of the quantum efficiency}

\begin{figure}
  \begin{center}
\includegraphics*[width=12cm, bb=0 0 684 506]{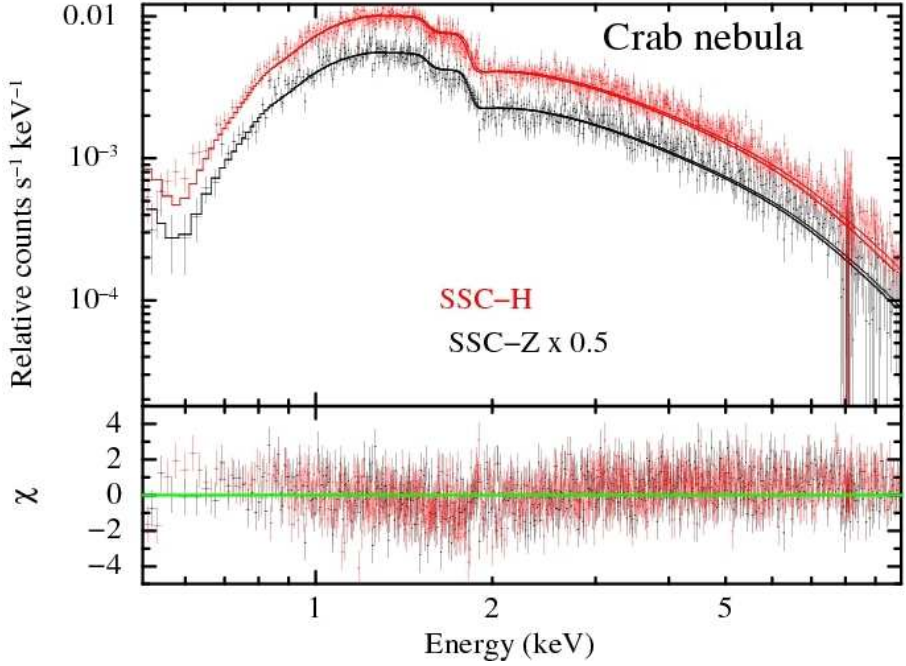}
\end{center}
  \caption{X-ray spectrum of the Crab nebula obtained by the SSC-H and the SSC-Z.  The solid lines show the best fit model (power law + N$_{\rm H}$).}\label{Crab_ssch_mid}  
\end{figure}

We accumulated data for the Crab nebula, shown in figure \ref{Crab_ssch_mid}.  When we fit the data by a power-law with interstellar absorption feature, N$_{\rm H}$, and the quantum efficiency (QE) of the pre-launch data (\cite{matsuoka2009}), we obtained the power law index $\gamma$ of 2.2 and N$_{\rm H}$ of 6$\times 10^{21}$cm$^{-2}$.  These are different values from those obtained in the previous literature (\cite{kirsch2005}).  With taking into account the energy range of the SSC, we find the parameters for the Crab nebula as $\gamma$ is 2.1 of N$_{\rm H}$ of 3.8$\times 10^{21}$cm$^{-2}$.  This indicates that we need further calibration.

\begin{table}[hbt]
\begin{center}
\caption{\sc Spectrum fit for the Crab nebula} 
\label{CCD_QEparameter}
\begin{tabular}{ccc} \hline\hline
Layer  &  design value   &  derived value  \\
\hline
Depletion layer [Si] & 70\,$\mu$m    &   75$\pm$ 3\,$\mu$m \\
Gate [Si] & 0.1\,$\mu$m    &   0.39$\pm$ 0.02\,$\mu$m \\
Insulator [SiO$_2$] & 0.8\,$\mu$m    &   0.79$\pm$ 0.04\,$\mu$m \\
Optical block [Al] & 0.2\,$\mu$m    &   0.21$\pm$ 0.01\,$\mu$m \\
\hline
\end{tabular}
\end{center}
\end{table}

The CCD employed has a Si depletion layer above which there are SiO$_2$ insulator, Si gate and Al coat.  We leave those parameters free so that we can reproduce the Crab nebula spectrum mentioned above.  The best fit parameters are shown in table \ref{CCD_QEparameter} as well as the design values.  We noticed that the thickness of Si shows the biggest difference from the design value.  Since the CCD chip is the coldest part in the MAXI, we may have contamination on it.  The difference may come from the possible contamination that will be checked in future calibration.

\begin{figure}
  \begin{center}
\includegraphics*[width=8cm, bb=0 0 643 495]{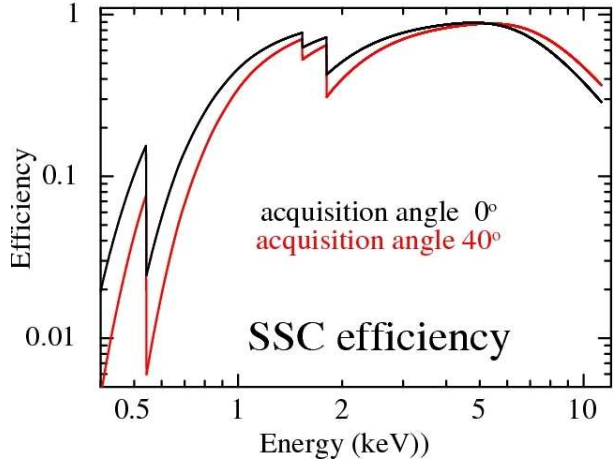}
\end{center}
  \caption{QE of the SSC calibrated by the Crab nebula spectrum.  We employ G0 for E$\le$1.8\,keV and G0+G1+G2 for E$\ge$1.8\,keV.}\label{QE_modified}  
\end{figure}

The spectrum obtained is the integration of all the data.  The MAXI/SSC detected the Crab nebula with various acquisition angle (0$^\circ \sim 40^\circ$).  The best fit parameters are converted to the observation of the acquisition angle at 0$^\circ$.  In the real data analysis , we need to modify the QE taking into account the acquisition angle.  Figure \ref{QE_modified} shows the revised QE for the MAXI/SSC.

%% file: ScanObservation.tex
\section{Scan Observation}

\begin{figure}
  \begin{center}
\includegraphics*[width=6cm, bb=0 0 593 592]{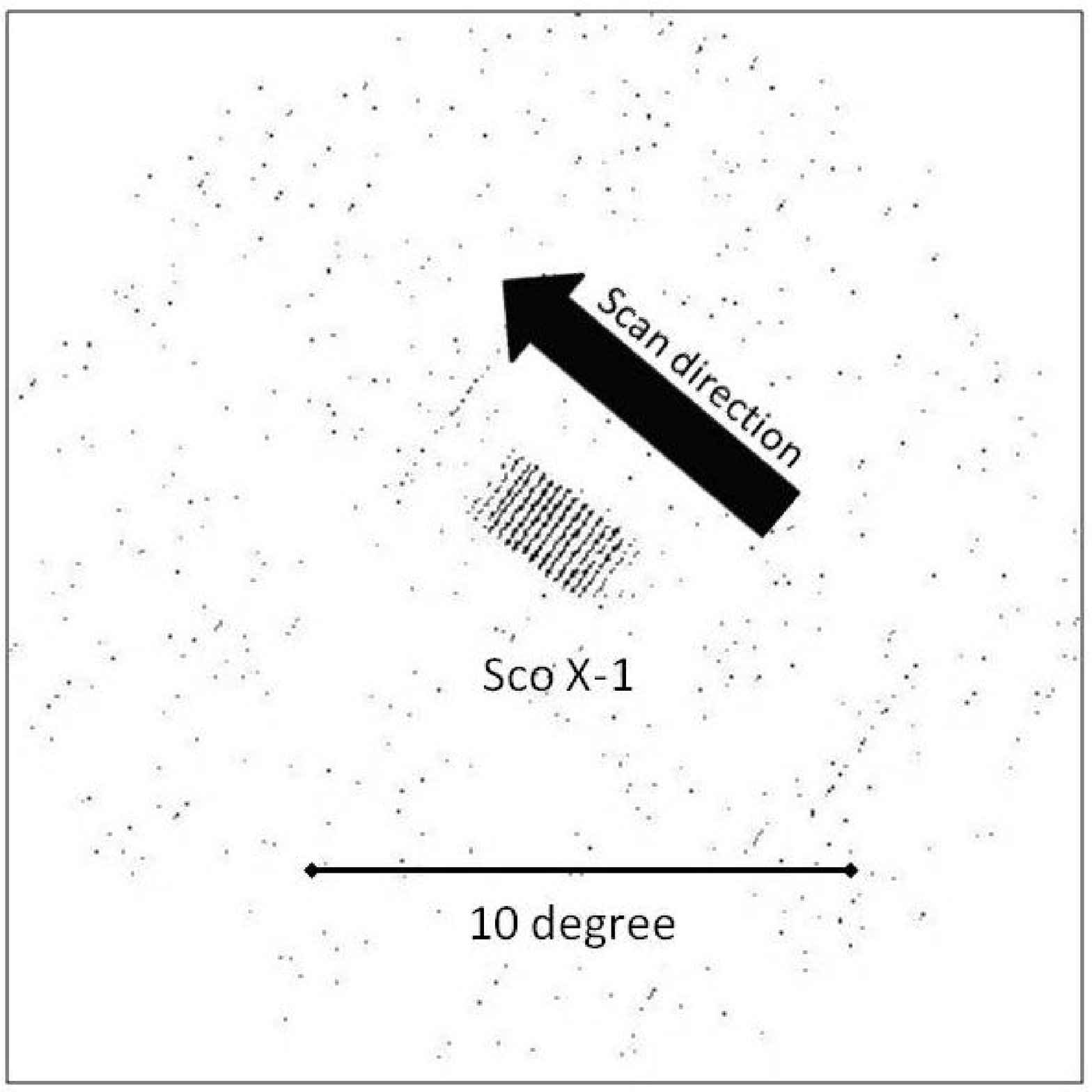}
\includegraphics*[width=6cm, bb=0 0 622 622]{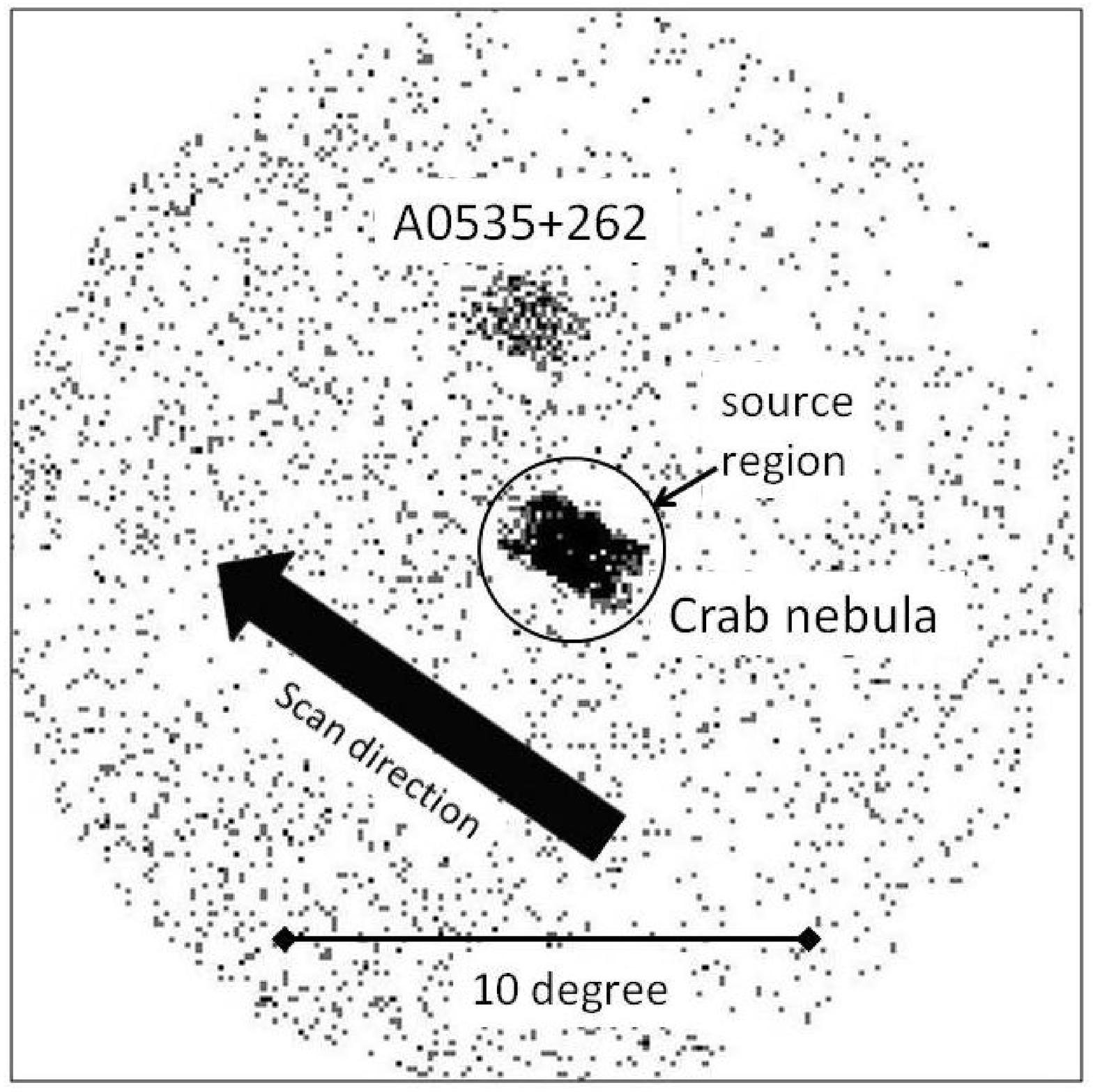}
\end{center}
  \caption{Singe scan image of the Sco X-1 (left) and one-day integration data of the Crab nebula (right).   Since the read-out time is \Rcycle\, s, the data along the scan direction is quantized in the left figure.}\label{one_scan}
\end{figure}

MAXI/SSC observes the sky with an FOV of $1^\circ.5 \times 90^\circ$ in each camera.  The angular response along the scanning direction is set by the collimator, resulting in a trianglar shape.  The perpendicular direction is limited by the slit.  Since the position resolution is limited by the CCD pixel size, it shows a box-car shape.  The FOV moves with the ISS rotation around the Earth.  Therefore, the on-source time is about 45\,s.  Figure \ref{one_scan} shows the distribution of photons on the sky when the Sco X-1 and the Crab nebula pass the FOV.   The figure for the Sco X-1 comes from a single scan.  The source is generally detected by two CCD chips out of 16 chips.  Therefore, the data are taken two times in one read-out cycle (\Rcycle\,s).  We see several line segments that are separated by about 2.9\,s.

The figure for the Crab nebula comes from the data of one-day integration where we also see a recurrent nova of A0535+26 (\cite{sugizaki2009}).  We see that the Crab nebula gives us 80 counts/scan.  The SSC can detect 1300\,photons/day from the Crab nebula if it is in the acquisition angle of 0$^\circ$.  The background level of the SSC is about 150\,photons/day/PSF.  We find that the Crab nebula is detected at the 100$\sigma$ confidence level in one-day integration data.  Therefore, the detection limit of the SSC is about 50mCrab (5$\sigma$) for a one-day observation.  It becomes about 200mCrab for single scans.

The CCD chip of the SSC can detect about 200\,photons\,cm$^{-2}$\,(read-out time)$^{-1}$ if we set the pile up events to be less than 10\%.  In the normal operation mode, this value corresponds to 80\,photons\,cm$^{-2}$\,s$^{-1}$ that is 25 Crab nebula intensity.  Therefore, we can expect that the source like the Sco X-1 is almost pile-up free even at the center of the FOV.

%% file: AllSkyMap.tex
\section{All Sky Map} 

\begin{figure}
  \begin{center}
\includegraphics*[width=12cm, bb=0 0 869 476]{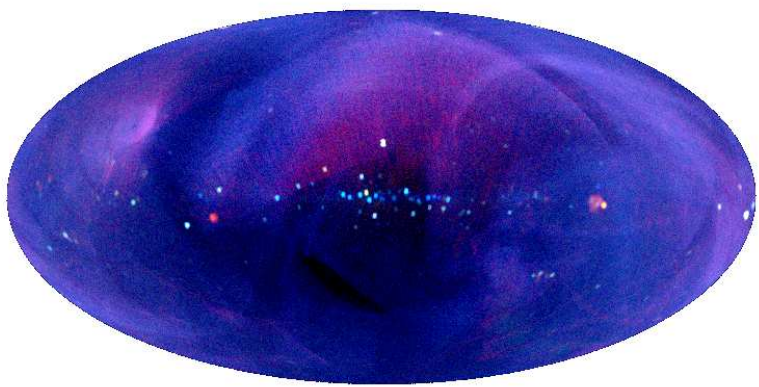}
\end{center}
  \caption{All sky image obtained by the SSC for 7 months.}\label{AllSkyImage}
\end{figure}

The SSC observes the sky all the time.  Only at the passage of the SAA and at the time when the Sun is very close to the FOV, we switch off the bias voltage applied to the CCD.  The data are  subject to the edge glow problem during day time observation.  Therefore, the effective observation efficiency is expected to be about \SSCeff \%.  In the first 6 months after the launch, we had occasional problems in one of the two data transfer systems.  When we had a problem, we had to turn the MAXI off to recover.  This reduced the data taking efficiency down to 20\% or less.  The problem was finally fixed by adding an extra router by an astronaut that was transported with the space shuttle in February, 2010.

In this way, we accumulated the data and obtained the all sky map by using 7 months data shown in figure \ref{AllSkyImage}.  The red, green and blue on the map correspond to the energy range of 0.6-1.0keV, 1.0-3.0keV and 3.0-8.0\,keV, respectively.  This map is exposure-time corrected.  It should be compared with that obtained by the GSC (\cite{mihara2010}).

Since the PSF of the SSC is similar to that of the GSC, the image in figure \ref{AllSkyImage} is similar to that of the GSC, however, the SSC is sensitive to the energy range below 2\,keV that can not be reached by the GSC.  The brightest sources below 2\,keV are the Cygnus Loop and the Vela SNR.  The Vela SNR is also seen its extent about 7$^\circ$ in diameter.

In the all sky map, we notice an extended structure, particularly extended in the northern hemisphere.  There are two problems in the SSC observation.  One is that the observation efficiency is about \SSCeff \%\ due to the edge glow effect and the passage of high background regions.  The other is the observation mode of the MAXI.  The MAXI is observing the sky all the time.  We have no chance to see the dark Earth from which we can estimate the non-X-ray background.  We have the possibility that the ISS is up-side down when the space shuttle is approaching to dock.  So far, we have insufficient observation time of the dark Earth which makes us difficult to detect emissions from extended sources.  Further observation and study are needed to obtain an extended emission analysis extraction that will be reported elsewhere.

%% file: Conclusion.tex
\section{Conclusion}

MAXI was launched in July, 2009 and installed on the ISS.  Observation started from August, 2010.  The CCD on the SSC is cooled down to the working temperature around $-60^\circ$C by the combination of the peltier cooler, LHP and radiator.  The CCD is working either in a frame mode or in a parallel sum mode (64-binning).  The frame mode is employed in the diagnostic mode.  The parallel sum mode is employed in the standard observation mode.  The detection limit (5$\sigma$ level) of the SSC is 200\,mCrab for one scan and 50\,mCrab for one-day integration.

The frame mode shows the performance of the CCD.  In the night time, the CCD functions properly while it suffers an edge glow when the Sun is near the FOV.  The observation efficiency of the SSC is about \SSCeff \%.  The performance of the CCD is continuously monitored both by the Mn-K X-rays and by the Cu-K X-rays.  In Mn-K X-rays, the energy resolution of the CCD is 147\,eV (FWHM) at the time of launch.  The radiation damage decreases the performance of the CCD.  Since we employ a charge injection method, the degradation of the CCD reduces to some extent.  The gain of the CCD stays almost constant while we measured the gain decrease to be 0.9\% /year without CI.  The FWHM decreases about 60\,eV/year.  The recovery by the CI of the SSC is not so effective to that of the XIS on Suzaku.  Both satellites are in low Earth orbit with different inclination angle.  The SSC passes high background region at high latitude that the XIS does not reach.  Therefore, the difference in the CI may be due to the particle background condition.

The SSC observes the all sky by using the X-ray CCDs for the first time.  There are many sources detected not only point sources but extended sources.  Due to the lack of effective observation time, we need more observation time to obtain an extended emission analysis extraction process.
